\documentstyle[11pt,AATS,epsf]{article}	
 
\begin{document}	
  
\title{Stellar Seismology, Stellar Ages and the Cosmological Constant}
   \author{G.R. Isaak}
   \affil{School of Physics and Astronomy, University of Birmingham, UK}
   \author{K.G. Isaak}
   \affil{Cavendish Laboratory, University of Cambridge, UK}
  
   \begin{abstract}
Solar seismology has allowed precision measurements of both the static and
dynamic structure of our local star, the Sun.
In the near future, seismology of solar-like stars of different ages and
masses, necessarily restricted by angular resolution to low l-modes, will
allow studies of the internal structure of stars at various stages
of evolution. Such studies will test not only the theory of
stellar evolution, but also allow the determination of ages of
stars from the helium content in their cores. Such observations can
be made photometrically from space, but also spectroscopically from the
ground. We outline ground-based schemes. By correlating the external 
properties of nearby stars with their internal properties,
it will be possible to extend local studies to distant open and
globular clusters, and thereby to obtain an age of the Universe, based on
many stars. The combination of the age, the density parameter
$\Omega$ and Hubble's constant will allow strong limits to be placed on the
cosmological constant.
\end{abstract}
  
  
   \section{Introduction}
  
 Recent measurements of distant, type Ia supernovae, taken as standard
  candles (Perlmutter {\it et al.}, 1999; Riess {\it et al.}, 1998), have 
  been interpreted as implying an accelerating expansion of the Universe 
  and demanding the existence of a cosmological constant and, presumably, 
  a new fundamental interaction (Zlatev, Wang \& Steinhardt, 1999). 
  The implication of such interpretations is profound,
  not just for cosmology, but for physics at large. Clearly, such
  extraordinary claims require extraordinarily good evidence, preferably
  by as many, independent means as possible.
  A possible method of verification is the determination of the ages of stars
  in our own galaxy, at $z=0$, and, thereby, a lower limit to the age of the
  Universe. With a Hubble constant $H_o=67$km s$^{-1}$ Mpc$^{-1}$  
  and a density parameter $\Omega=1$, the age of the Universe is 9.6 Ga 
  if the cosmological constant, $\Lambda=0$. 
  Adopting the currently fashionable cosmological vacuum energy 
  contribution of $\Omega_{\Lambda}=0.7$ demands an age of 13.8 Ga.
 
  To measure stellar ages requires a calibration of the scale on which 
  such age determinations are based -- a calibration of the theory of 
  stellar evolution, which has usually been taken as sacrosanct. 
  The Sun provides a calibration, albeit at one point only in the parameter
  space of mass, composition and age. An external calibration, using 
  only the readily measured external parameters has, of course, been used
  in the past. An internal calibration, using parameters of the solar 
  (stellar) core, is also, in principle, possible. This provides a 
   near-direct measure of the 
  molecular weight in the core and, thereby, measure of
  the helium content.
  If the original amount of helium is presumed to be known and if 
  gravitational settling of helium is allowed for, the current helium content
  provides a measure of the amount of thermonuclear conversion of hydrogen 
  into helium and therefore, with a known luminosity, the age of the star.

Stellar seismology, an extension of the successful solar seismology, is 
capable of providing this information both in principle and, in the near 
future, in practice for stars over a range of ages. There is prospect that 
a reasonable zero-age calibration point, in addition to that given by the 
Sun, could be provided by a very young star, preferably one with known 
mass and composition. The absolute error in the theoretically determined 
age is then likely to be sufficiently small. 
This, via a cosmological aeon ladder, ought to allow us to determine 
the ages of old stars and also of those in old open and globular clusters.

We propose to use the existing seismological calibration of the Sun 
to bootstrap and thereby determine the ages of stars, and thus check the 
theory of stellar evolution over a wide range of stellar masses, 
internal composition and importantly age. Such an undertaking, using 
stellar seismology, is feasible using current technology and is underpinned
by extensive work that has gone into establishing the acoustic eigenmode 
spectrum of the Sun over the last two decades. 

Here, we outline, albeit very sketchily, the relevant key elements of solar 
seismology, its extension to stars, the sensitivity to age of the 
relevant eigenfrequency separations and the feasibility of those 
measurements from a space platform as well as from the ground.

\begin{figure}[t!]
\plotfiddle{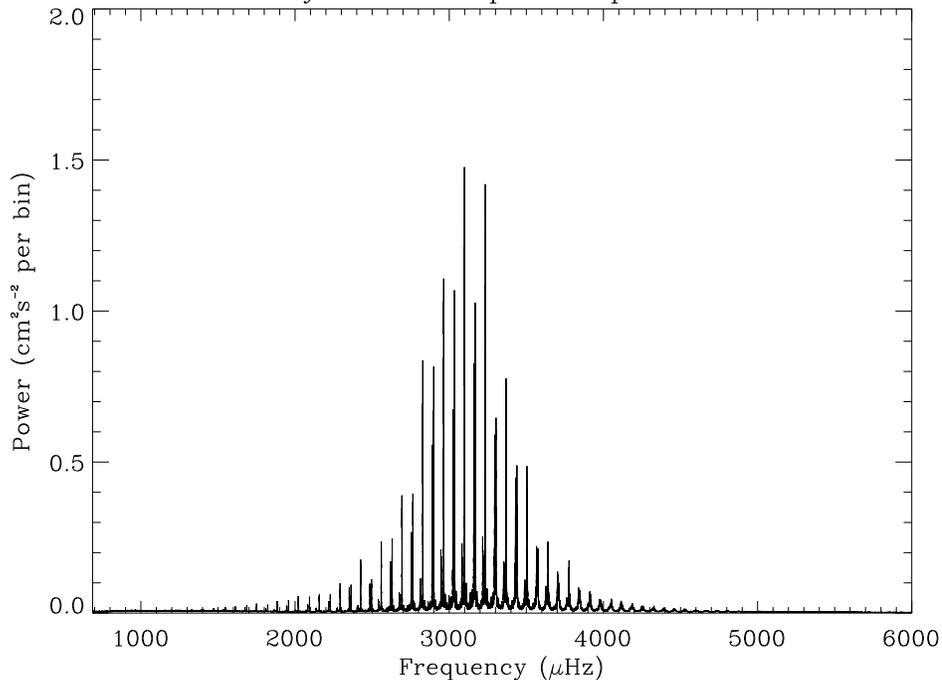}{8cm}{90}{55}{55}{200}{-50}
\caption{ A power spectrum generated from 7 years of data accummulated from
6 stations of BiSON, the Birmingham Solar Oscillation Network (Chaplin 
{\it et al.}, 1996 ) 
\label{ref_sunpowerspectrum} }
\end{figure}

\section{Solar Seismology}
Electromagnetic spectroscopy of the eigenmodes of the atom has provided a 
wealth of tools with which to probe Nature at large over the last one and 
a half centuries. It would seem that the detailed study of the acoustic 
spectrum of a star, (see Figure \ref{ref_sunpowerspectrum}), and in 
particular the study of eigenfrequencies, shows equal promise. 
Unlike many other parameters pertaining to stellar systems,
eigenfrequencies are both precise {\it and} accurate. Frequency 
measurements with an accuracy of up to a few parts per million, relevant in our 
context, have been made with no systematic error. 
Some statistical errors are, however, present: in particular, those 
arising from the finite duration of any  observation, the finite 
lifetime of the eigenmode itself, daily aliases 
and additional amplitude noise. Inspite of these, it was possible 
already back in 1981 (Claverie {\it  et al.}, 1981) to determine the accoustic 
eigen-frequencies of the individual low $\it{ l,  l=0,1,2,3}$ modes 
of the global Sun ( Sun as star) to better than 1 part in 4000. 
Currently, the accuracy attained is better than 1 in $10^5$ 
(Chaplin {\it et al.}, 1998). 
 
Individual eigenfrequencies provide integral measures of the
velocity of sound, and thus of $\sqrt{\gamma kT /m}$ over
  the ray path of the sound wave, with different modes probing 
  different depths.  At present the accuracy to which the eigenfrequencies 
  can be measured far outstrips the level to which theoretical models and  
  observational data agree. Thus, little useful science can, at present, 
  be extracted from direct comparisons between observed eigenfrequencies 
  and their 'postdiction' by theoretical models. A large part of this 
  problem is believed to be the incomplete treatment of the contributions
  of the surface layers to the travel time of sound waves. 
  By taking {\it differences} between eigenfrequencies this problem
  can be substantially eliminated. Differential information gleaned by 
  measuring frequency separations includes: 
  (a) the separation between successive harmonics of the radial and 
   low {\it l } 
  non-radial modes - this large separation, $\Delta$, provides a 
  measure of the time taken by  
  sound waves to traverse the diameter of the star -- 
 $\Delta\sim (2\int dr/c)^{-1}$.
 (b)  the small separation, $d_o$ between radial and quadrupole 
  $p$ modes -- closely spaced in frequency; these modes share a very similar
  path in the outer regions of the Sun, but 
  penetrate to different depths in the core, and thus measure 
  core properties.
 (c)  an even smaller splitting due to the removal of degeneracy of the
 non-radial eigenmodes by rotation, the rotational splitting: 
the number of components, 
$2{\it l  + 1}$, can be used to identify the mode, with their
spacing providing an integral measure of the internal rotation.

The use of such differencing methods is common in other branches
of physics. Perhaps the most familiar example can be found in an
atomic analogue: 
from fine-structure, hyperfine-structure and isotope shifts of observed 
optical atomic transitions it is possible to infer details about both 
atomic and nuclear properties that are very much more subtle than could 
be inferred from theoretical modelling of the atomic term values using 
Dirac's equation. In a similar way $d_o$ provides a sensitive measure of 
the physical conditions -  temperature and molecular weight -- in the 
core of the star. As the conditions vary with mass and age, so too does the 
separation -- in the Sun, its value is close to $9\mu$Hz. 
As a star evolves, hydrogen is converted into helium by 
thermonuclear fusion and, in the absence of mixing, the helium fraction in the 
nuclear active core increases. The resultant increase in the molecular 
weight in those regions sampled more by the radial than the quadrupole 
modes, results in a decrease in the velocity of sound and thus in the value of
$d_o$. The variation of both $d_o$ and $\Delta$ with evolutionary age for stars of different masses has been evaluated by J.Christensen-Dalsgaard 
(1984,1988) --  for a 
solar-type star at the midpoint of its life, as is the Sun, 
the sensitivity of the fine-spacing is near $1\mu$Hz/Ga.
Thus, the accuracies to which $\Delta$ and $d_{o}$ were measured in the 
Sun already back in 1981,  about $0.3\mu$Hz, correspond to an error in the 
determination of the solar age of 0.3 Ga. Current measurements have errors 
some twenty times smaller. Thus, it is 
clear that it is model-dependent, and NOT observational, error that will 
dominate any error-budget.
  
\section{Calibration and pitfalls}
The calibration of any cosmological aeon glass is full of pitfalls. 
Whilst the value for the age of the meteorite-Earth system, determined 
using radioactive decays of isotopes of $K$, $Rb$, $Sm$, $Lu$, $Re$, 
$Th$ and $U$, has
remained within $\pm 0.01$ Ga of the value of 4.55 Ga over the last 45 years 
(Dalrymple, 1994), the value for the age of the (oldest) globular clusters 
has varied over the range of 5 to 25 Ga over the same time interval 
(Sandage, 1962; Gamow, 1964).
This in itself serves as an indication that absolute calibration of the 
theory of the structure and evolution of stars is essential.
Such calibration of the basic physics has been provided by solar
seismology for the last twenty years, while theoretical solar models have been 
adjusted repeatedly over that time to approximate to the measured 
eigenfrequencies. The calibration of a model, solar or otherwise, 
at time {\it t}, means that we adjust the proposed model either {\bf
quantitatively} by adjusting parameters or {\bf qualitatively} by means 
of the addition of another physical mechanism such that model age and 
radioactive age agree exactly.
Two examples, however, may serve to demonstrate some of the pitfalls in 
such work. The first comes from the discovery measurements of global solar
oscillations (Claverie {\it et al.}, 1979). One of us, GRI (Isaak, 1980) 
used the extreme high frequency end of the eigenfrequencies of the solar models
of Iben \& Mahaffy (1976) and Christensen-Dalsgaard, Gough \& Morgan 
(1979), models which had been constructed in response to the apparently 
successful results of a long-term campaign to search for the 
fundamental radial and low n non-radial eigenmodes in 1974-5. The theoretical
models did not predict the excitation of high {\it n}, low {\it l} 
acoustic modes. It was fortuitous that the published values extended 
into the region in which, totally unexpectedly (Unno {\it et al.}, 1979), 
global solar oscillations were discovered. 
GRI deduced, by interpolation, that the best fit between $\Delta$ 
and the models implied a helium abundance of less 
than $Y=0.17$, in agreement with the average solar wind value {\it
but} 
the observational data, thought to be accurate to the 0.3\% level, 
were indeed correct, however the
model was wrong and required substantial quantitative adjustments.     
Thus, the potentially fundamental cosmological implication of this
very low Y inference was false. 

A second example is based on the much improved measurements of $d_o$ 
which BiSON (Birmingham Solar Oscillations Network) has been producing
consistently since the early 1990's. Once again,
GRI compared the measured values of $d_o$  with the much improved models
of Christensen-Dalsgaard, concluding that the fit was poor -- assigning 
an age of 5.2 Ga to the Sun would have provided an excellent match between 
model and data,  however this would have been totally inconsistent with the
radioactively-determined solar age, which by then
was very secure. In principle, the Sun could be older than the rest
of the Solar System, however, the presence of daughter products of short-lived 
isotopes such as $^{26}Al$, $^{107}Pd$ and $^{41}Ca$ in meteorites 
suggests strongly that the meteorite formation was coeval with the Sun 
(Guenther, 1989, Bahcall {\it et al.},1995), 
presumably triggered by a supernova 
explosion. Models suggest that it takes around 50 Ma for a molecular 
cloud to collapse onto the zero age main sequence (ZAMS) 
(Schwarzschild 1958, Iben 1965). Thus, the age of the Sun, measured from 
its arrival on the ZAMS is 4.56-0.05=4.51 Ga, with an uncertainty 
which would seem to be at most 0.02 Ga. GRI preferred the radioactive 
age to the seismological age and speculated, in common with others
(Christensen-Dalsgaard, Proffit \& Thompson, 1993), that helium settled 
gravitationally (Isaak, 1993), with the idea shortly afterwards that heavier
elements (Proffitt, 1994) settled also. These {\it qualitatively
new} 
additions to the physics of the seismological Sun can very nearly, but 
not perfectly, reproduce the measurements 
(Elsworth {\it et al.}, 1995, Chaplin {\it et al.}, 1997).   
A cynical reader might wonder as to the relevance of the above: 
clearly, the measurements are of an integral nature and no unique 
interpretation is possible. We suggest, however, that the
study of solar seismology has substantially enriched the theory of 
stellar structure and evolution, beyond that first proposed by 
Eddington of 1928, by confronting theory with accurate observation.
The future of stellar seismology has even greater potential, if 
appropriate financial resources were to be made available.

\section{Can one detect stellar oscillations?}
Soon after the discovery of global solar oscillations 
(Claverie {\it et al.}, 1979),
it was pointed out by GRI (Isaak 1980) that the detection of 
stellar oscillations of the size seen on the Sun was readily achieved 
in one of two ways: 
   (a) by measuring spectroscopic velocities using large flux collectors 
(b) by photometric measurements using a photometer and very modest 
    optical telescopes situated above the transparency fluctuations and 
    the turbulence of the terrestrial atmosphere. 

Two years later, GRI also suggested a stellar seismology mission to the 
European Space Agency; however neither this mission, nor any of its 
many successors over the last 18 years, have been launched.
  
Can stellar oscillations really be detected? Here, we assess the 
feasibility of the detection of solar-like  modes in main-sequence stars. 
We assume (a) oscillation amplitudes of a size comparable to those seen on the
Sun -- there is some evidence to suggest (Houdek {\it et al.}, 1995)
 that the more evolved 
a star, the larger the amplitude of oscillation. We adopt, however, a 
conservative amplitude of $\sim 10 cm/s$ in velocity, and 
2 parts per million in intensity (b) an extension of spectroscopic 
measurements from the Sun to 
the brightest local stars is difficult as the fluxes of the 
Sun : Sirius : 3.3 $^m$ star are  
in the ratio 1:$10^{-10}$:$10^{-12}$.
  
To determine $\Delta$ and $d_o$ it is necessary to measure the 
stellar eigenmodes, and to resolve them in order to be able to 
measure them sufficiently accurately. This requires that each of 
three distinct, and fundamental (and physically-based) observational 
criteria are met: 
   (a) that each mode is measured with a signal-to-noise ratio, $S/N >4$:
   many claims of the detection of stellar modes have been made over the
   the last 17 years, largely with low S/N data and complex data analysis. 
   Not one detection, to date has, however,  been substantiated. 
   (b) that the uncertainty principle is satisfied -- if we assume
   that the oscillations have a lifetime greater than the observation time, 
   then to measure with an error of $1(0.3)\mu Hz$, with a commensurate
   error in stellar age of 1 (0.3)Ga, requires 2 (6) weeks. 
   (c) that the Nyquist sampling theorem is satisfied -- at least 2 
    measurements need to be made in the shortest time period (highest 
    eigenfrequency) of interest. 
  
Each of the above need to be met for observations that are shorter than 
the lifetime (coherence time) of the mode. 
If the observations are longer than the coherence time of the mode, then 
the power spectra of observations separated by more than the coherence 
time are independent and gain of accuracy with time is slow. If incoherent, 
the frequency errors in the power spectrum scale only as the square root 
of time.
  
As an additional constraint, it is clear that the instrumental stability 
and any resultant systematic errors must not be larger than the 
statistical contribution to the noise levels.
  
\subsection{Photometry from Space}
To achieve the required photometric accuracy of 2 parts per million,
roughly equivalent to a velocity amplitude of $10cms^{-1}$ (Isaak, 1980), 
is extremely difficult, or even impossible, from ground level sites. Such 
accuracy, however, can be readily achieved from a platform above most, 
or all, of the atmosphere. Whilst an expensive solution, the feasibility 
of such photometry was amply confirmed when solar oscillations were 
detected in 1982 at the predicted 
level (Isaak 1980) with ACRIM (Active Cavity Radiometer) on the 
Solar Maximum Mission (Woodard \& Hudson 1983). 
A minimum of $4\rm{x}10^{12}$ photons has to be detected, and 
systematic errors have to be correspondingly small. A telescope of 
$1 (0.3)m$ diameter collects such an integrated flux in just 
over 1 (c. 20)  day(s)  from an 8th magnitude star.
  
\subsection{Ground-based Spectroscopy}
Stars subtend angles of less than some 10 milliarcsec, a factor of 
$10^{-5}$ of the Sun. Thus, effects of differential extinction across 
the stellar disk due to the terrestrial atmosphere, a source of 
significant systematic error in measurements of global solar oscillations, 
are negligible. By using differential measurement methods and 
rapid switching, spectroscopic techniques can be used to 
reduce systematic errors substantially, again as demonstrated by
work on the unresolved Sun over more than two decades. 
By switching between the blue and the red wings of a spectral 
line, 'common mode' effects are reduced substantially. Photon noise is,
however, a severe problem as the instantaneous fractional bandwidth of
spectroscopic instruments is usually small. 
In principle, we could consider three different scenarios of 
spectroscopic measurement: (a) a portion of a spectral line of bandwidth of 
50 $m\AA$   (b) all of one spectral line of $0.2 \AA$  width, for a star
that is rotating slowly (c) the measurement of a number, {\it n} ,
of spectral lines. Statistically, we gain a factor of two in going from 
(a) to (b), and by an additional factor of $\sqrt{n}$ by going to (c). 
Whether systematic errors can be kept down to correspondingly low values 
in (c) is unclear.
If we consider 
an ideal spectrometer, with a throughput of unity, with bandwidth
as specified by case (a), fed by a 3 (10) metre-diameter flux collector, 
a total of $7\rm{x}10^4$ ($7\rm{x}10^5$) photons per second would be
collected from a star of 3.3$^m$.  
The latter counting rate is within a factor of two of the Mk I 
optical resonance 
   scattering spectrometer viewing the Sun (Brookes {\it et al.}, 1978) 
of BiSON on 
   Tenerife and was one of the two spectrometers separated spatially by 
   some 2300 km, with which global solar oscillations were first discovered 
   in 1979 (Claverie {\it et al.}, 1979). 
Clearly, such a flux collector - ideal 
   spectrometer of type (a), i.e. even of 
   $0.05\AA$ bandwidth, can readily detect stellar oscillations with typical 
   solar amplitudes. \newline
We propose to discuss another specific spectroscopic
   system - the magneto-optical-filter (MOF). The MOF was developed by Oehman
   (1956) and by Cimino, Cacciani \& Sopranzi (1968)
   . An improved version of the MOF (Isaak \& Jones, 1986) 
   has been used repeatedly and has reached completely photon 
   noise limited velocity noise
   at the $35cms^{-1}$ level (Bedford {\it et al.}, 1995). 
  The throughput of that spectrometer 
   is down on the ideal by about 400. Use of two polarisations (gain: *2), 
   a CCD as a detector rather than an avalanche photodiode (gain: *4), and 
   the use of NaD1 and NaD2, as well as the potassium resonance line at 
   $7699\AA$ restores a factor of 24. The solar spectral line has a slope 
   which is a factor of *4 steeper. 
   The overall expected performance, scaling from the ACTUAL PERFORMANCE, 
   with a gain of $4\rm{x}\sqrt{24}$
 improvement on the Bedford {\it et al.} figure would be 
   a velocity noise level of $2cms^{-1}$ on a $0.3^m$ star with two weeks 
   observing time on a 1.9 m diameter telescope, or $8cms^{-1}$ for 
a $3.3^m$ star.
   Clearly, flux collectors of the Hanbury-Brown type with modern electronic
   active optics, and diameters of  6 metres would make the above possible.
   It should be stressed that the optical resonance spectrometers have an
   enormous etendue and can, therefore, accept beams from poor quality, i.e.
   cheaply made, flux collectors.
 
\section{Summary}  
\begin{itemize}
\item Global solar oscillations probe a $1M\odot$ Pop-I star at 
$4.51$ Ga, and so {\it calibrate} the aeon glass at one point in time,
mass and composition. Another point, the zero-age point,
may be provided by stars which are theoretically reckoned to be extremely 
young.
\item Stellar oscillations of main-sequence stars could calibrate
aeon glasses over a range of ages.
\item Both photometry from space and spectroscopy from the ground
provide sufficient sensitivity to render the detection of stellar
oscillations feasible.
\item By measuring the large and the small spacings of the acoustic 
eigenmode spectrum of old solar-like stars their ages could be 
measured to substantially better than 1 Ga; these 'internal' measurements 
could then be used to calibrate the 'external' characteristics of the 
same stars. Transferring this calibration, using 'external' 
characteristics to more distant stars, population I as well as II, 
then allows the ages of old open and globular clusters to be determined.
\item Together with the present value of $H_o$=67 kms$^{-1}$Mpc$^{-1}$, 
$\Omega=1$, stellar ages can be used to check on the existence of a 
cosmological constant $\Lambda$. $\Lambda=0$ demands an age of the 
Universe of 9.6 Ga, whereas the currently fashionable $\Omega_{\Lambda}=0.7$
requires an age of the Universe of 13.8 Ga. Given that any stellar ages 
provide at best a lower bound to the age of the Universe, if any star were
found to have an age in excess of 10 Ga, then stellar seismology would 
provide an independent confirmation of the need for a non-zero cosmological 
constant. Stellar ages below 10 Ga would, in contrast, provide necessary
but not sufficient evidence to exclude a finite $\Lambda$ if the oldest stars
in our galaxy were to be considerably younger than the Universe. 
 
\item We stress that in contrast to the use of SNIa and other techniques, 
seismological measurements can thus be considered to make possible 
cosmology studies at zero redshift.
\end{itemize}


  
   \end{document}